\documentclass[twoside,fleqn]{ActaStyle}
\usepackage{times,cite}

\def\authorheading{Davor Palle}
\def\shorttitle{On gravitational lensing ...}

\begin{document}

\pagerange{1}{8}

\title{On gravitational lensing by quadrupole potentials}

\author{Davor Palle \email{palle@mefisto.irb.hr}}
{Zavod za teorijsku fiziku, Institut Rugjer Bo\v skovi\' c \\
Po\v st. Pret. 180, HR-10002 Zagreb, CROATIA}

\day{March 23, 2005}

\abstract{
We study gravitational lensing by quadrupole potentials
within the linearized gravity approximation  and the integration 
over the unperturbed photon trajectory.
It is well known that the quadrupole potential contribution 
to the deviation angle is much smaller than that of the 
monopole one. We show, however, that quadrupole potentials
can change the photon polarization vector, but
there is no contribution from the monopole term to the first
order. 
The effect is maximal when the axis of the quadrupole is
tangential to the photon trajectory and it is 
proportional to the frequency (rate) of the quadrupole
moment of the deflector.
The second order correction of the monopole potential
to the polarization is canceled away by the
renormalization of the polarization vector.
%PACS numbers may be entered using the \verb+\pacs{#1}+ command.
}


\section{Introduction and motivation}

The common belief in physics is that Einstein's general theory
of relativity inevitably implies the appearance of 
gravitational waves.
The mathematical structure of these waves is defined as
quadrupole radiative fields.

However, one can question about two issues relevant 
to gravity and gravitational waves:
(A) The existence problem: to prove the existence of
gravitational waves, it is necessary to derive the exact
wave equation from general relativity, but this is
impossible because of the presence of at least
Newtonian "monopole" terms that spoil the structure
of the "wave" equation \cite{synge,weinberg,landau,misner};
any approximate or even exact solution of  Einstein
field equations cannot isolate or neglect Newtonian 
from quadrupole terms.
(B) The locality problem: assuming the presence of
gravitational waves infers that a decomposition 
of the total tensor field contains terms affecting
and implying the curvature of spacetime, essentially the nonlocal
physical process, and at the same time the local radiative quadrupole
field as gravitational waves; such a decomposition is
difficult to comprehend both physically and mathematically.

The classical electrodynamics was the inspiration to
introduce gravitational waves, but (a) the dipole
radiative field is a solution of the exact Maxwell
wave equations and (b) the electrodynamics is
completely defined only by the local gauge field.
As a consequence, the predicted dipole radiation has been
verified experimentally. 

It is possible to understand the energy loss of
binary systems as the effect of quadrupole
potentials and not as a result of radiated
gravitational waves 
\cite{palle}.

In this paper we inspect gravitational lensing
by quadrupole potentials which are parts of
the metric tensor, such as  monopole terms, 
affecting spacetime curvature.
It should not be confused with the lensing of
the assumed
gravitational radiation by cosmological environment or
some astrophysical sources that one can find in the 
literature.

\section{Lensing equations}

Gravitational lensing by quadrupoles is an old subject
of investigation \cite{zipoy,winterberg,bertotti,trevese}, but the first
reliable calculation is due to
Damour and Esposito-Far\`{e}se \cite{damour}.
They have shown that the incorrect results of Ref. \cite{durrer} arise
from a naive "plane wave" approximation.
Unfortunately, the same unreliable approximation is used in 
Ref. \cite{kopeikin}.

Let us briefly summarize the short-wave approximation
of Maxwell equations in curved spacetime and
the transport along the rays of the wave vector ($\ell^{\mu}$),
the scalar amplitude (a) and the complex polarization vector ($V^{\nu}$) 
\cite{schneider}:

\begin{equation}
\ell^{\mu}\equiv \frac{d x^{\mu}}{d \xi},\ \ell^{\beta}\ell^{\alpha}_{;\beta}=0,
\nonumber \\
\dot{V_{\alpha}}\equiv \ell^{\beta}V_{\alpha ;\beta}=0,\ 
\dot{a}+\frac{1}{2}a \ell^{\alpha}_{;\alpha} = 0, \nonumber
\end{equation}
\begin{eqnarray*}
\eta_{\mu \nu}=diag(-1,+1,+1,+1),\ V^{*}_{\nu}V^{\nu}=1,\ 
\ell^{\alpha}\ell_{\alpha}=0,  \\
\dot{a}\equiv \ell^{\beta}a_{,\beta},\ V_{\alpha}\ell^{\alpha}=0
\end{eqnarray*}

We can write the transport equations more explicitly using
Christoffel symbols:

\begin{eqnarray}
\frac{d \ell_{\mu}}{d \xi} &=& \frac{1}{2}\ell^{\alpha}\ell^{\beta}
\partial _{\mu}g_{\alpha\beta}, \\
\frac{1}{a}\frac{d a}{d \xi} &=& -\frac{1}{2}(\partial_{\beta}\ell^{\beta}
+\Gamma^{\alpha}_{\alpha \lambda}\ell^{\lambda}), \\
\frac{d V_{\mu}}{d \xi} &=& \ell^{\alpha}V_{\beta}\Gamma^{\beta}_{\mu \alpha}.
\end{eqnarray}

The linearized gravity approximation and the integration
over the unperturbed photon trajectory \cite{schneider}
of the transport equations, lead to the following lensing equations:

\begin{eqnarray*}
g_{\mu \nu} = \eta_{\mu \nu} + h_{\mu \nu},\ b^{\alpha}\equiv impact\ vector,\ 
b^{\mu}\ell_{\mu}=0, \nonumber 
\end{eqnarray*}
\begin{equation}
\Delta \ell_{\mu} \equiv \ell_{\mu} (out) - \ell_{\mu} (in) =
\frac{1}{2}\ell^{\alpha}\ell^{\beta} \int_{-\infty}^{+\infty} d\xi 
\partial_{\mu} h_{\alpha \beta} (b^{\lambda} + \xi \ell^{\lambda})\ , 
\end{equation}
\begin{equation}
\Delta ln\ a \equiv ln \frac{a(out)}{a(in)} = -\frac{1}{4}\ell^{\lambda}
\int_{-\infty}^{+\infty} d\xi \partial_{\lambda} h (b^{\mu} +
 \xi \ell^{\mu}),\ h\equiv \eta^{\alpha\beta}h_{\alpha\beta}, 
\end{equation}
\begin{equation}
\Delta V_{\alpha} \equiv V_{\alpha}(out)-V_{\alpha}(in)=
\frac{1}{2}\ell^{\beta}V^{\gamma}\int^{+\infty}_{-\infty}
d\xi (\partial_{\alpha}h_{\beta\gamma}+\partial_{\beta}h_{\alpha\gamma}
-\partial_{\gamma}h_{\alpha\beta})(b^{\mu}+\xi \ell^{\mu}).
\end{equation}

A direct evaluation of the lensing equations with the monopole potential
straightforwardly gives

\begin{eqnarray}
h^{(1)}_{\alpha\beta}(x^{\mu}) = \frac{2 G_{N} M}{\mid \vec{x} \mid}\delta_
{\alpha\beta} \Longrightarrow 
\Delta \ell_{\mu} = -\frac{4 G_{N} M}{b^{2}} b_{\mu},\ 
\Delta ln\ a = 0,\ t^{\mu}\Delta V_{\mu} = 0, \\
t^{\mu}=any\ spacelike\ vector\ orthogonal\ to\ \ell^{\mu}. \nonumber
\end{eqnarray}

Thus, besides the resulting 
nonvanishing Einstein deflection angle \cite{schneider},
the monopole potential does not influence the transport of the scalar amplitude or
of the polarization vector to the first order.

For the calculus with quadrupole potentials, we adopt
the formalism of Damour and Esposito-Far\`{e}se \cite{damour}
performing the integration in the Fourier space.
They evaluated the lensing of the wave vector, which we therefore omit and
concentrate on the lensing of the scalar amplitude and the polarization
vector.

One can read the Fourier transform of the potentials where 
the energy-momentum conservation is also included \cite{damour}:

\begin{eqnarray*}
h_{\mu \nu} (x^{\lambda}) = \int \frac{d^{4} k}{(2\pi)^{4}}
\widehat{h}_{\mu \nu} (k^{\lambda}) e^{i k\cdot x}  , 
k\cdot x \equiv k^{\nu} x_{\nu}, \widehat{T}\equiv \eta^{\mu\nu}\widehat{T}_{\mu\nu}, 
\end{eqnarray*}
\begin{equation}
\widehat{h}_{\mu \nu} (k^{\lambda}) = 16 \pi G_{N} \frac{\widehat{T}_{\mu
\nu} (k^{\lambda}) - \frac{1}{2} \eta_{\mu \nu} \widehat{T} (k^{\lambda}) }{k\cdot k - 
i \epsilon k^0} \ , \\
\widehat{T}_{0i} = - \frac{k^j}{k^0} \widehat{T}_{ij},\ 
\widehat{T}_{00} =  \frac{k^i
k^j}{(k^0)^2} \widehat{T}_{ij} .
\end{equation}

We can fix the constant wave ($\ell^{\mu}$), impact ($b^{\nu}$) and
polarization ($V^{\alpha}$) vectors:

\begin{equation}
\ell^{\mu}=(1,0,0,1),\ b^{\mu}=(0,b,0,0),\ V^{\mu}=(0,V_{1},V_{2},0) .
\end{equation}

There is no influence of the quadrupole as a deflector to the scalar amplitude:

\begin{equation}
\Delta ln\ a (quadrupole) = 0.
\end{equation}

Changing the integration variable $k\equiv\mid \vec{k} \mid$ to $u=\sqrt{k^{2}-\omega^{2}}$ 
\cite{damour},
one obtains the expression for the lensing of polarization:

\begin{eqnarray*}
k^{\mu}\equiv(\omega,k\sin\vartheta\cos\phi,k\sin\vartheta\sin\phi,
k\cos\vartheta),\ k\cdot V\equiv k^{\nu}V_{\nu}, 
\end{eqnarray*}
\begin{eqnarray}
\Delta V_{\alpha} =  \frac{i G_{N}}{\pi^{2}}\int^{+\infty}_{-\infty}
d \omega \int^{+\infty}_{0}d u \int^{2\pi}_{0} d \phi  
 u e^{i bu\cos \phi} (u^2-i \varepsilon \omega)^{-1} \nonumber \\
\times \{k_{\alpha}\widehat{T}_{ij}(-k^{j}V_{i}\omega^{-1}+\delta_{i3}V_{j}) 
-(k\cdot V)\widehat{T}_{ij}[\delta_{\alpha 0}(k^{i}k^{j}\omega^{-2}-
\delta_{i3}k^{j}\omega^{-1})+\delta_{\alpha i}(-k^{j}\omega^{-1}+\delta_{j3})]
\nonumber \\
+\frac{1}{2}(k\cdot V)\ell_{\alpha} (-k^{i}k^{j}\omega^{-2}\widehat{T}_{ij}
+\widehat{T}_{ii})\}.
\end{eqnarray}

Neglecting the $\vec{k}$ dependence (quadrupole approximation) of the deflector field
\cite{damour}

\begin{eqnarray}
\widehat{T}_{ij}(\omega,\vec{k})\simeq \widehat{T}_{ij}(\omega,\vec{0})
=-\frac{\omega^{2}}{2}D_{ij}(\omega), \ \ \ \ \ \ \ \ \ \ \\
D_{ij}(\omega)\equiv \int d^{3}x x^{i}x^{j}T^{00}(\omega,\vec{x}),\ 
D_{ij}(\omega)\equiv \int dt e^{i \omega t}D_{ij}(t), \nonumber
\end{eqnarray}

the rest of integrations could be performed with elementary integrals but
with a careful regularization procedure \cite{damour}.

The result for the lensing of polarization is of the form

\begin{equation}
\Delta V_{1}(Q) = \frac{4 G_{N} V_{2}}{b^{2}}\frac{\partial D_{12}(t)}
{\partial t}\mid_{t=0},\ 
\Delta V_{2}(Q) = -\frac{4 G_{N} V_{1}}{b^{2}}\frac{\partial D_{12}(t)}
{\partial t}\mid_{t=0},
\end{equation}
\begin{eqnarray*}
\Delta V_{0}(Q) &=& -\frac{2 G_{N}}{b^{3}}[V_{1}(D_{22}(0)-D_{11}(0)
+\frac{b^{2}}{2}\frac{\partial^{2}D_{11}(t)}{\partial t^{2}}\mid_{t=0}
-\frac{b^{2}}{2}\frac{\partial^{2}D_{22}(t)}{\partial t^{2}}\mid_{t=0}) \\
&+& V_{2}(2D_{12}(0)+b^{2}\frac{\partial^{2}D_{12}(t)}{\partial t^{2}}
\mid_{t=0})],
\end{eqnarray*}
\begin{eqnarray*}
\Delta V_{3}(Q) &=& -\frac{2 G_{N}}{b^{3}}[V_{1}(D_{22}(0)-D_{11}(0)
-\frac{b^{2}}{2}\frac{\partial^{2}D_{11}(t)}{\partial t^{2}}\mid_{t=0}
+\frac{b^{2}}{2}\frac{\partial^{2}D_{22}(t)}{\partial t^{2}}\mid_{t=0}) \\
&+& V_{2}(2D_{12}(0)-b^{2}\frac{\partial^{2}D_{12}(t)}{\partial t^{2}}
\mid_{t=0})].
\end{eqnarray*}

One can easily verify the following
 relation

\begin{eqnarray*}
V_{\mu}^{*}\Delta V^{\mu}(Q)+V^{\mu}\Delta V_{\mu}^{*}(Q)=0.
\end{eqnarray*}

From the exact integral for the evolution along 
the geodesic (C) of the polarization vector
\begin{equation}
\Delta V_{\mu} = \int _{C}d \xi \ell^{\alpha}V_{\beta}
\Gamma ^{\beta}_{\mu\alpha},
\end{equation}
one can easily deduce the second order corrections
perturbing Christoffel symbols \cite{weinberg}, wave and polarization
vectors, as well as the photon trajectory with regard
to the monopole potential:

\begin{equation}
\Delta V_{\alpha}(M)={\cal I}_{\alpha}+{\cal J}_{\alpha}
+{\cal K}_{\alpha}+{\cal L}_{\alpha}+{\cal M}_{\alpha},
\end{equation}

\begin{eqnarray}
{\cal I}_{\alpha}&=&\frac{1}{2}\ell^{\beta}V_{\lambda}
\int_{C_{0}}d \xi h^{(1)\lambda\kappa}(\partial_{\alpha}
h^{(1)}_{\beta\kappa}+\partial_{\beta} h^{(1)}_{\alpha\kappa}
-\partial_{\kappa} h^{(1)}_{\alpha\beta}), \\
h^{(1)}_{\mu\nu}&=&\frac{2 G_{N}M}{\mid \vec{x}\mid}
\delta_{\mu\nu},\ h^{(1)\kappa\gamma}=-\eta^{\beta\gamma}
\eta^{\alpha\kappa}h^{(1)}_{\alpha\beta},\ 
C_{0}=unperturbed\ trajectory, \nonumber
\end{eqnarray}

\begin{eqnarray}
{\cal J}_{\alpha}&=&\frac{1}{2}\ell^{\beta}V^{\kappa}
\int_{C_{0}}d \xi (\partial_{\alpha} h^{(2)}_{\beta\kappa}
+\partial_{\beta} h^{(2)}_{\alpha\kappa}-
\partial_{\kappa} h^{(2)}_{\alpha\beta}), \\
h_{00}^{(2)}&=&-\frac{2 G_{N}^{2}M^{2}}{\mid \vec{x}\mid^{2}},\ 
h_{ij}^{(2)}=\frac{G^{2}_{N}M^{2}}{\mid \vec{x}\mid^{2}}
(\delta_{ij}+\frac{x_{i}x_{j}}{\mid \vec{x}\mid^{2}}),\ 
h^{(2)}_{i0}=0, \nonumber
\end{eqnarray}

\begin{eqnarray}
{\cal K}_{\alpha}=\frac{1}{2}V^{\kappa}\int_{C_{0}}
d \xi \Delta \ell^{\beta}(\partial_{\alpha}h^{(1)}_{\beta\kappa}
+\partial_{\beta}h^{(1)}_{\alpha\kappa}
-\partial_{\kappa}h^{(1)}_{\alpha\beta}), \\
\Delta \ell_{\alpha}(\xi)=\frac{1}{2}\ell^{\mu}\ell^{\nu}
\int^{\xi}_{-\infty}d \xi \partial_{\alpha}
h^{(1)}_{\mu\nu}(b^{\lambda}+\xi \ell^{\lambda}), \nonumber \\
\Longrightarrow \Delta \ell_{\alpha}(\xi)
=G_{N}(0,\frac{-2 M \xi}{b\sqrt{b^{2}+\xi^{2}}}-\frac{2 M}{b},
0,\frac{2 M}{\sqrt{b^{2}+\xi^{2}}}), \nonumber 
\end{eqnarray}

\begin{eqnarray}
{\cal L}_{\alpha}&=&\frac{1}{2}\ell^{\beta}\int_{C_{0}}
d \xi \Delta V^{\kappa}(\partial_{\alpha}h^{(1)}_{\beta\kappa}
+\partial_{\beta}h^{(1)}_{\alpha\kappa}-
\partial_{\kappa}h^{(1)}_{\alpha\beta}),  \\
\Delta V_{\lambda}&=&\frac{1}{2}\ell^{\beta}V^{\kappa}
\int^{\xi}_{-\infty}d \xi (\partial_{\lambda}h^{(1)}_{\beta\kappa}
+\partial_{\beta}h^{(1)}_{\lambda\kappa}
-\partial_{\kappa}h^{(1)}_{\lambda\beta})(b^{\nu}+\xi \ell^{\nu}), 
\nonumber 
\end{eqnarray}
\begin{eqnarray*}
\Rightarrow \Delta V_{\lambda} =G_{N}(\frac{M V_{1}}{b}(1+\xi /
\sqrt{b^{2}+\xi^{2}}),\frac{M V_{1}}{\sqrt{b^{2}+\xi^{2}}},
\frac{M V_{2}}{\sqrt{b^{2}+\xi^{2}}},
\frac{M V_{1}}{b}(1+\xi/\sqrt{b^{2}+\xi^{2}})), 
\end{eqnarray*}

\begin{eqnarray}
{\cal M}_{\alpha}&=&\frac{1}{2}\ell^{\beta}V^{\kappa}
\int_{C_{1}}d \xi (\partial_{\alpha}h^{(1)}_{\beta\kappa}
+\partial_{\beta}h^{(1)}_{\alpha\kappa}
-\partial_{\kappa}h^{(1)}_{\alpha\beta}), \\
C_{1}&=&perturbed\ trajectory, \nonumber \\
\Rightarrow {\cal M}_{\alpha}&=&\frac{1}{2}\ell^{\beta}V^{\kappa}
\int_{C_{0}}d \xi \Delta x^{\rho}\frac{\partial}
{\partial x^{\rho}}[\partial_{\alpha}h^{(1)}_{\beta\kappa}
+\partial_{\beta}h^{(1)}_{\alpha\kappa}
-\partial_{\kappa}h^{(1)}_{\alpha\beta}], \nonumber \\
\Delta x_{\alpha}&=&\frac{1}{2}\ell^{\beta}\int^{\xi}_{-\infty}
d \xi x^{\kappa}(\partial_{\alpha}h^{(1)}_{\beta\kappa}
+\partial_{\beta}h^{(1)}_{\alpha\kappa}
-\partial_{\kappa}h^{(1)}_{\alpha\beta}), \nonumber 
\end{eqnarray}
\begin{eqnarray*}
\Rightarrow \Delta x_{\alpha}=G_{N}(M(1+\xi/\sqrt{b^{2}+\xi^{2}}),
\frac{3bM}{\sqrt{b^{2}+\xi^{2}}},0,3M(1+\xi/\sqrt{b^{2}+\xi^{2}})
-2M \ln\frac{\xi+\sqrt{b^{2}+\xi^{2}}}{\epsilon}),  \\
\epsilon=small\ positive\ real\ cut-off. 
\end{eqnarray*}

The coordinate singularity in Eq.(21), regulated by 
the cut-off $\epsilon$, disappears after the integration
over the unperturbed photon trajectory, as one should
expect for an observable.
After performing integrations, only one term contributes
to the deviation of the polarization vector:

\begin{eqnarray}
{\cal I}_{\alpha}&=&G_{N}^{2}(-\frac{2\pi M^{2}}{b^{2}}V_{1},0,0,
-\frac{2\pi M^{2}}{b^{2}}V_{1}), \nonumber \\
{\cal J}_{\alpha}&=&G_{N}^{2}(-\frac{\pi M^{2}}{2b^{2}}V_{1},0,0,
\frac{3\pi M^{2}}{8b^{2}}V_{1}), \nonumber \\
{\cal K}_{\alpha}&=&G_{N}^{2}(0,\frac{4 M^{2}}{b^{2}}V_{1},
\frac{4 M^{2}}{b^{2}}V_{2},\frac{2\pi M^{2}}{b^{2}}V_{1}), 
\nonumber \\
{\cal L}_{\alpha}&=&G_{N}^{2}(\frac{3\pi M^{2}}{2b^{2}}V_{1},0,0,
\frac{\pi M^{2}}{2 b^{2}}V_{1}), \nonumber \\
{\cal M}_{\alpha}&=&G_{N}^{2}(-\frac{2\pi M^{2}}{b^{2}}V_{1},0,0,
-\frac{2\pi M^{2}}{b^{2}}V_{1}), \nonumber \\
&\Rightarrow &\Delta V_{1}(M)=\frac{4 G_{N}^{2}M^{2}}{b^{2}}V_{1},\ 
\Delta V_{2}(M)=\frac{4 G_{N}^{2}M^{2}}{b^{2}}V_{2}. 
\end{eqnarray}

However, this contribution is canceled away by the
renormalization of the polarization vector:

\begin{eqnarray*}
\Delta V_{\mu}&\equiv &\Delta V_{\mu}(Q)+\Delta V_{\mu}(M),\  
(V^{\mu}+\Delta V^{\mu})^{*}(V_{\mu}+\Delta V_{\mu})=1
+\frac{8 G_{N}^{2}M^{2}}{b^{2}}+O((\Delta V)^{2}), \\
\tilde{V}^{\mu}&\equiv &(1+\frac{8 G_{N}^{2}M^{2}}{b^{2}})^{-\frac{1}{2}}
(V^{\mu}+\Delta V^{\mu}),\ \tilde{V}^{\mu}\tilde{V}_{\mu}=1, \\
\tilde{V}^{\mu}&=&V^{\mu}-\frac{4 G_{N}^{2}M^{2}}{b^{2}}V^{\mu}
+\Delta V^{\mu}+O((\Delta V)^{2})
=V^{\mu}+\Delta V^{\mu}(Q)+O((\Delta V)^{2}),\ \mu=1,2.
\end{eqnarray*}

\section{Results and discussion}

For example, take a binary system and its quadrupole moment
when a surface of motion is in the xy plane and the z axis is defined by the direction
of the incoming photon. In this case, the deflection of 
polarization is maximal. Generally, the Euler rotation matrix R
projects the quadrupole moment of arbitrary orientation to the
frame defined by the wave vector of the deflected photon

\begin{eqnarray}
\tilde{D}_{ij}(t) = \frac{m_{1}m_{2}}{m_{1}+m_{2}}d^{2}
\left[ \begin{array}{ccc}
\cos^{2}(\Omega t) & \frac{1}{2}\sin (2\Omega t) & 0 \\
\frac{1}{2}\sin (2\Omega t) & \sin^{2}(\Omega t) & 0 \\
0 & 0 & 0 \end{array} \right ], \nonumber \\
\Omega\equiv \frac{2\pi}{P},\ P=orbital\ period, \nonumber \\
circular\ orbit:\ x=d\cos (\Omega t),\ y=d\sin (\Omega t), \nonumber
\nonumber \\
D_{ij}(t) = R(\theta_{E},\phi_{E})_{3\times 3} \tilde{D}_{ij}(t).\ \ \ \ \ \ \ \ \ 
\ \ \ \ 
\end{eqnarray}

The complex polarization vector and its deflections could be cast
into a more conventional form to recognize linear,
circular, or elliptical polarization states 
(here $\beta$ is the phase-difference between two 
independent polarization states) \cite{jackson}

\begin{eqnarray}
\vec{V}=V_{1}\vec{x}_{0}+V_{2}\vec{y}_{0}=
V_{-}\vec{\epsilon}_{+}+V_{+}\vec{\epsilon}_{-},\ \ \ \ \ \ \ \ \  \nonumber \\
V_{\pm}=\frac{1}{\sqrt{2}}(V_{1}\pm i V_{2}),\ 
\mid V_{-} \mid ^{2} + \mid V_{+} \mid ^{2} = 1,\ 
\vec{\epsilon}_{\pm} = \frac{1}{\sqrt{2}}(\vec{x}_{0}
\pm i \vec{y}_{0}), \nonumber \\
V_{+}=\rho_{+},\ V_{-}=\rho_{-}e^{i \beta},\ 
\hat{\rho}=\frac{\rho_{-}}{\rho_{+}}=\frac{\sqrt{1-\rho_{+}^{2}}}
{\rho_{+}}, \nonumber \\
\Longrightarrow \Re V_{1}=\frac{1}{\sqrt{2}}(\rho_{+}+
\rho_{-}\cos \beta),\ \Im V_{1}=\frac{1}{\sqrt{2}}\rho_{-}\sin \beta,
\nonumber \\
\Re V_{2} = -\frac{1}{\sqrt{2}}\rho_{-}\sin\beta,\ 
\Im V_{2} = \frac{1}{\sqrt{2}}(\rho_{-}\cos \beta - \rho_{+}),
\nonumber \\
\Delta \rho_{+} = \frac{1}{\sqrt{2}}(\Re \Delta V_{1}
-\Im \Delta V_{2}), \ \ \ \ \ \ \ \ \ \ \  \\
\Delta \beta = \frac{1}{\sqrt{2}\rho_{-}\cos \beta}
(\Im \Delta V_{1}-\Re \Delta V_{2}+
\sqrt{2}\sin \beta \frac{\rho_{+}}{\rho_{-}} \Delta \rho_{+}), \\
quadrupole\ contribution(Eq.(14)): 
\Delta \beta = \frac{4 G_{N}}{b^{2}}\frac{\partial D_{12}(t)}
{\partial t}\mid_{t=0},\ 
\Delta \rho_{+}=\Delta \hat{\rho} = 0 . \ \ \ \ \ 
\end{eqnarray}

Finally, let us present some numerical estimates of the effect
assuming high-frequency (high-rate) deflectors:

\begin{eqnarray*}
binary\ neutron\ star\ 
 system:m=m_{1}m_{2}/(m_{1}+m_{2}),
 \ \dot{D}_{12}(0) = md^{2}\Omega,
\ \ \ \ \  \\
Kepler's\ third\ law:\ d=\Omega^{-2/3}(G_{N}(m_{1}+m_{2}))^{1/3},  \\
m\simeq M_{\odot},\ d\simeq 10^{-4} s,\ \Omega\simeq 10^{3}s^{-1},\ b\simeq 10^{-3}s
\Longrightarrow  \Delta \beta = O(10^{-4}),
\ \ \ \ \ \ \ \ 
\end{eqnarray*}

and similarly, for the accretion by supermassive black holes or
coalescing black hole binaries. 

Although the effect is small, the deviation of polarization 
(change of the phase-difference between the two independent states)
by the quadrupole is free of any "background" monopole contribution
and it is worth measuring.

\end{document}